# Near-infrared scintillation of liquid argon: recent results obtained with the NIR facility at Fermilab

C. O. Escobar[a*], P. Rubinov[a] and E. Tilly[b]

[a] *Fermi National Accelerator Laboratory*
*Kirk Rd and Pine St., Batavia, IL 60510, U.S.A*

[b] *University of the South*
*735University Ave., Sewanee, TN 37383-1000, U.S.A*

E-mail: escobar@fnal.gov

ABSTRACT: After a short review of previous attempts to observe and measure the near-infrared scintillation in liquid argon, we present new results obtained with NIR, a dedicated cryostat at the Fermilab Proton Assembly Building (PAB). The new results give confidence that the near-infrared light can be used as the much needed light signal in large liquid argon time projection chambers.

KEYWORDS: Time projection chambers; near-infrared scintillation; dark matter detectors.

---

[*] Corresponding author.

**Contents**

## 1. Introduction

   In recent years the technology of liquid noble gases (LNG) detectors [1] has come of age and is finding widespread use in experiments searching for dark matter (DM) [2] and in large liquid argon (LAr) time projection chambers (TPC) for neutrino physics [3]. In both cases (DM and neutrino physics) there are significant gains when detecting both the charge coming from the primary ionization as well as the scintillation light produced in an event through excitation mechanisms in the LNG medium. The primary gain for neutrino physics is to determine the initial time for the drifting electrons for non-beam related events such as supernova neutrinos and searching for nucleon decay, while for DM detectors the light signal provides an extra handle for the discrimination between nuclear and electron recoils.

   Regarding the scintillation light, the focus of this work, so far, all experiments running or planned to run in the immediate future, consider detecting the abundant vacuum-ultraviolet (VUV) scintillation light from noble gases either in gaseous or condensed states with wavelengths as short as 78 nm for liquid neon up to 175 nm in liquid xenon. The challenges presented by the detection of the short wavelength VUV light are well known and will not be repeated here but suffice to say that all experiments up to now need to use wavelength shifters (WLS) such as tetra phenyl butadiene (TPB) [4] or p-terphenyl (pTp) [5] to shift the VUV light to wavelengths suitable for the collection, transport and final detection by currently existing and affordable photosensors [6]. Questions arising from the long term stability of surfaces coated with TPB [7] or pTp [5] immersed in LNG have been raised from time to time in the literature and are now becoming an overpresent concern by many groups [8].

   Significant Rayleigh scattering, when propagating in the LNG media, is another unavoidable feature of the VUV light. This is not a concern for small volumes of LNG but indeed becomes a source of performance degradation as the volume increases and light coming from as far as 2.5 m away from the photon collection system needs to be efficiently detected [9]. A recent theoretical estimate of the Rayleigh scattering length [10] places it at 55 cm a value that seems to be supported by indirect measurements by two different DM experiments, ArDM and DarkSide-50 [11].

   The difficulties outlined above are not a comprehensive list of the challenges introduced by the use of WLS in LNG detectors due to the insistence in detecting the VUV light. For the last couple of years we have been proposing to investigate in a systematic way if LNGs have a significant emission in the NIR and if this light could be used either as a replacement or in addition to the VUV signal [12]. If a significant light yield (LY) is found in the near-infrared portion os the spectrum it could then be envisaged a detection scheme avoiding the use of wavelength shifters with the extra benefit of curtailing the Rayleigh scattering. The efficient detection of NIR light in a way that can be used in the design and production of an affordable photon detection system for DM or neutrino detectors remains to be demonstrated but here and in previous works [12] we describe the first steps towards this goal.

## 2. Review of past results

   The only possibility of turning the NIR light from condensed noble gases into a useful signal in large volume detectors is if such a emission comes not from atomic transitions but rather from self-trapped excitons in much the same manner as the VUV light. It has been known that near-infrared emission from noble gases occurs with emission lines around 1300 nm [13] but this emission is expected to be quenched in the condensed phase of the noble element in question. In the late seventies, early eighties, investigators using the technique of transient optical absorption spectroscopy via electron excitation, observed NIR transitions in noble gases both in the gaseous state [14] and in condensed form [15]. The spectral features of the NIR transitions in the condensed phases when compared to the transitions in gas away from the atomic lines, led to the interpretation that these features are due to the formation of self-trapped excitons, raising the expectation that this process is not quenched nor the NIR light absorbed as it propagates in the condensed noble gas medium. The main spectral feature in liquid argon, observed in ref. [15], has a broad peak around 1.27 eV with a long tail extending down to 1.1 eV, ie from 976 nm down to 1130 nm and up to 1.4 eV or 890 nm. It is important to stress the broadness of this spectrum as the short wavelength part overlaps with the sensitivy region of the photon detectors that were used in our studies, as will become clear when describing the experimental setup.

### 2.1 More recent results

In the current decade two research groups have been very active in the investigation of the NIR scintillation in LAr, one at the Technical University of Munich, Germany and the other from the Budker Institute in Novosibirsk, Russia. Representative and early publications by these groups can be found in refs [16] and [17], respectively. The experimental approach of these two groups have in common the fact that they both use table-top setups, with small volumes of LAr, of the order of a few cubic centimeters and that both groups excite the liquid argon through the use of very intense, low-energy, pulsed beams, a pulsed source of X-rays with energies between 30 and 40 keV for the Russian group and an electron gun (pulsed and continuous) of 12 keV in the case of the Munich group. It is worth emphasizing that our approach, to be described in the next section, uses excitation mechanisms more akin to the sources of ionizing radiation in DM and neutrino detectors.

Since the results from both groups have been previously reviewed in this conference series (see [12] and [18]), we will just present a quick summary of their findings pointing out their relevance for our continued investigation of the subject.

The initial results from the Munich group [16][19] indicated a continuum for wavelengths longer than 700 nm with a broad peak at around 970 nm, close to the cut-off wavelength of their spectrometer, leading the authors to state that "liquid argon might have some rather strong emission features in the near-infrared". This result was followed up by the Novosibirsky
group [20, 21] which, lacking a spectromer to resolve the wavelengths of the scintillation, took the unnormalized spectrum of Heindl et al. and folded it with photon detection efficiency (pde) of their infrared photosensitive silicon photonmultiplier (SiPM) with peak sensitivity at around 600 nm extending up to 1,000 nm (for the record, the pde at these long wavelengths are minimal, of the order of a few percent, see the manufacturer's data sheet [22]). The final ingredient in the calculation of the integrated light yield is the geometrical acceptance of their detection system which they obtain by Monte Carlo. The light yield is then obtained by matching the number of counted photoelectrons in the SiPMs. The result is a light yield of $5.1 \times 10^2$ photons/MeV in LAr in the range 400 -1,000 nm.

The Munich group continued their studies now with an improved purification system added to their apparatus [23] and the possibility of adding Xe in a controlled way, with the result that they no longer see any relevant emission in pure liquid argon in the range 500 to 3,000 nm[23]. They therefore revise their previous result ([16],[19]) stating that they now believe the observed signal was an "artifact due to the normalization of the spectrum with the response function of the spectrometer used for that spectral range" [23].

.

## 3. The NIR facility at Fermilab and first results

The main components of the NIR facility at Fermilab were previously used in a project investigating the application of solid xenon as a detection medium and it was then known as the Solid Xenon Test Stand, described in detail in ref. [24]. In this section we will present only the main features of the facility that are needed for an understanding of the measurements and results presented here. The NIR cryostat, pictured in Figure 1, consists of a 30 cm external diameter stainless steel vacuum jacketed vessel with three 15 cm diameter observation glass windows. Inside this vessel there are two concentric Pyrex glass chambers, the outermost having a diameter of 23 cm and the innermost, with 10 cm diameter, being the one that contains the argon. This inner chamber has 5 mm thick side walls and a 10 mm thick flat bottom. The outer chamber is used as a cooling bath for the inner one and is filled with liquid nitrogen (LN). Argon from a dewar is flown through a hot getter for purification. Platinum resistance temperature detectors (RTD: PT-100) glued to the bottom and outer surface of the argon chamber monitor temperature gradients and a level probe monitors the level of LN in the outer chamber. The level of LAr in the inner chamber is controlled through the levels of LN in the cooling chamber. Once argon starts to condense in the inner chamber the LN level is lowered to prevent excessive cooling of argon which would lead to the formation of solid argon.

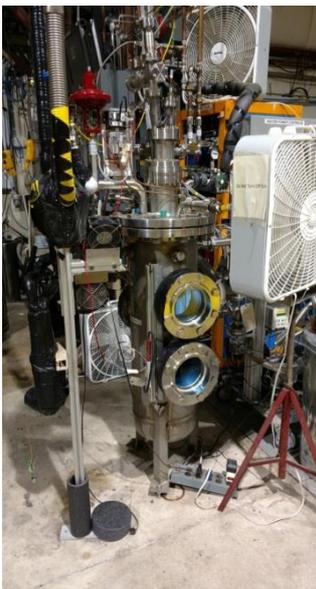
Figure 1. A picture of the NIR Cryostat showing two observation windows

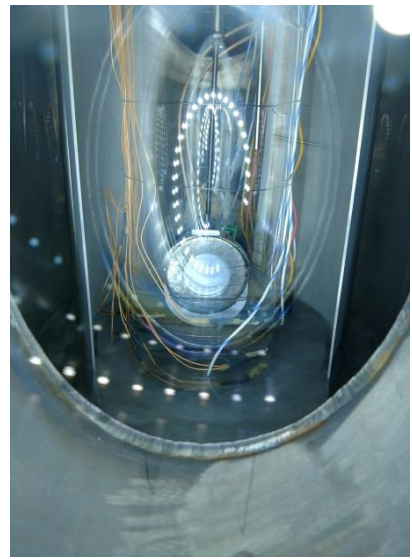
Figure 2. The two glass chambers and the Am-241 source in LAr.

In order to excite the Ar in the inner chamber we employ an Am-241 source of approximately 1 μC activity emitting alpha particles of 5.4 MeV. This source can be moved up and down inside the inner glass chamber being attached to the end of a rod that provides the excursion without breaking the sealing of the chamber from the external environment.

Figure 2 shows a picture of the interior of the cryostat with LN in the cooling chamber and a column of liquid argon in the inner chamber with the Am-241 source immersed in the liquid argon.

### 3.1 Experimental setup

A lay-out of the experimental setup is shown in Figure 3. As described in the previous section the Am-241 radioactive source is attached to a rod and the source can be moved up and down inside the argon chamber. Inside the argon chamber there are two NIR LEDs (850 nm peak λ) far from the cryostat window where an IR PMT (Hamamatsu R669) is attached with its window covered by a long pass infrared filter at 715 nm. Figs 4 and 5 show the filter transmission and the R669 spectral response, respectively. The quantum efficiency of the PMT at 715 nm is approximately 6% and, at 850 nm, 0.7%.

The PMT is operated at 1450 V and the signal is read by a digital scope. Trigger threshold is set at a level such that it can still trigger on dim light from the LED (dimming the light by decreasing the voltage on the LED as well as the pulse width). The trigger counts from the scope are sent to a 225 MHz universal counter which counts triggers on a 10 s interval for a couple of hours. We then perform a classical on-source versus off-source counting experiment, where on-source means that the source points out to the PMT and off-source indicating that the source is high up in the argon chamber at a level much above the cryostat window. Since the rod controlling the vertical position of the cryostat also allows for the rotation of the source around the rod axis, we also took data with the source facing away from the PMT but at the same level as the on-source position. For all practical purposes, the Am-241 source behaves like a point source of light given the short range of the 5 MeV alpha particles in liquid argon (a few micron). In both off-source positions some reflected infrared light will shine onto the PMT, and this constitutes an irreducible background in the experiment, with the facing away off-source position in principle producing more reflected light. This provides us with a good handle in assessing the background.

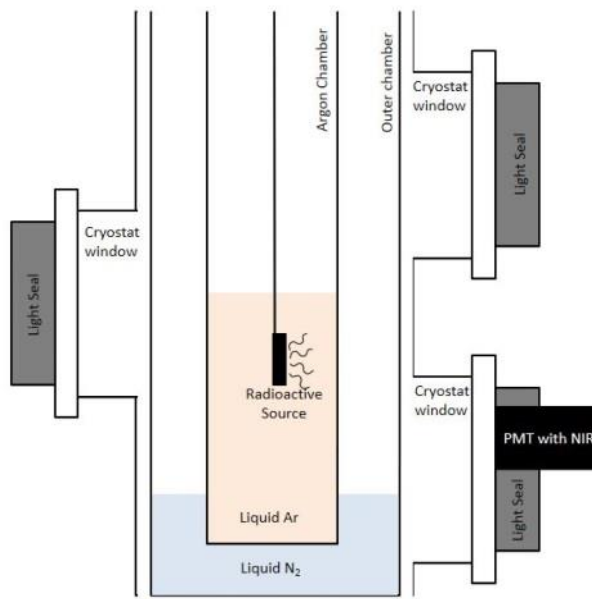

Figure 3. "Lay-out" of the experimental setup

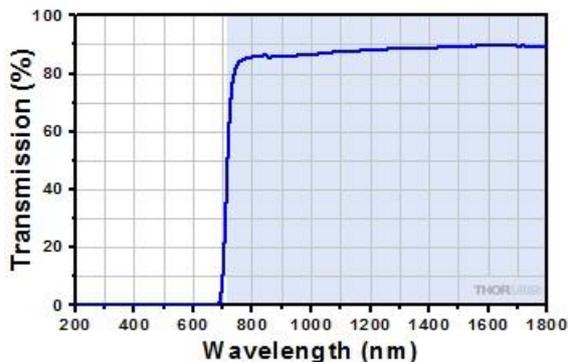

Figure 4. Transmission curve of the IR filter
Data from Thorlabs (www.thorlabs.com)

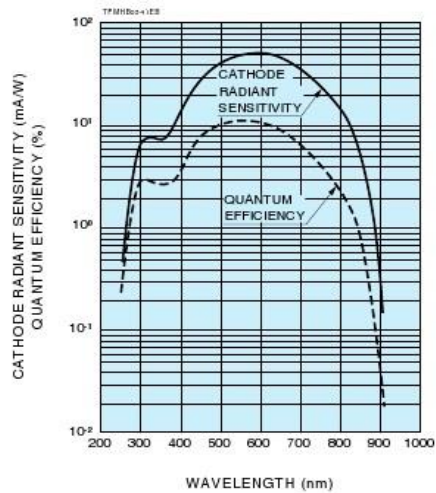

Figure 5. Spectral response of the IR PMT.
Data from Hamamatsu-R669.

### 3.2 First results

As described above the experimental setup is used to perform a counting experiment *on-source vs. off-source* where *on* means that the Am-241 source is at the same level as the center of the PMT window and facing it and *off* means either that the source is displaced all the way up inside the inner chamber, more than 40 cm above the PMT or that the source is at the same level as the PMT but facing away, rotated $180^0$. This last configuration is used as a handle in understanding the background since in this position more reflected NIR light is expected to reach the PMT as compared to the source all the way up.

We took data during several weeks in the months of July and August 2017 and a typical data set looks like the histograms show in Figure 6, where it can be clearly seen that there is a large separation in the counting rates for on-source and off-source configurations

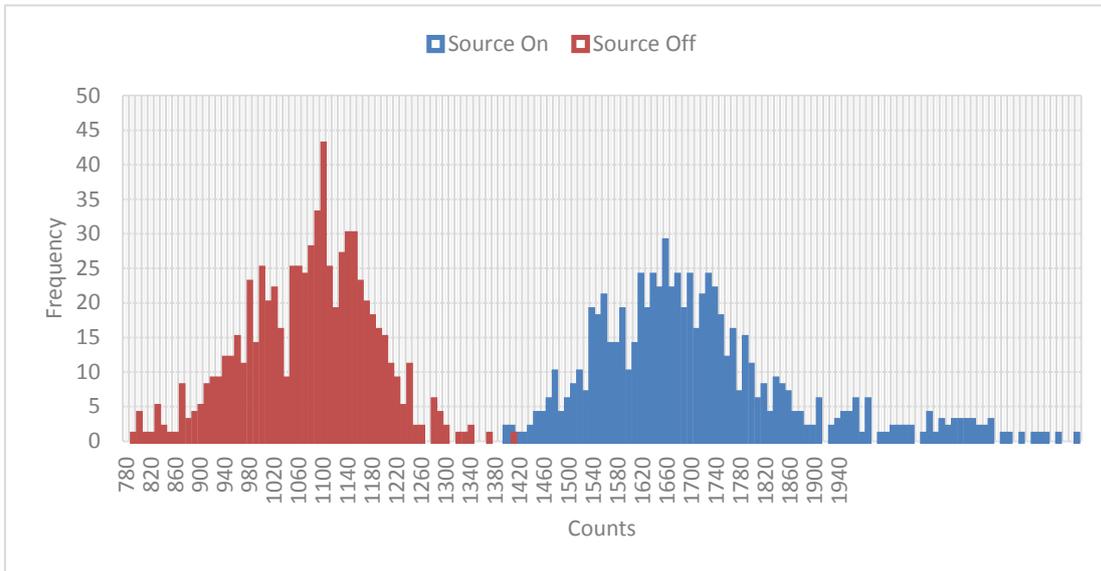

Figure 6. The distribution of 10 seconds counts for off-source (dark brown) and on-source (blue).

We perform a Kolmogorov-Smirnov test of these two data sets (on and off) with the null hypothesis being that the two data sets come from the same physical process. Figure 7 shows the cumulative distributions for the two data sets. The maximum difference is D = 0.564 corresponding to a probability P =0.000 ruling out the null hypothesis.

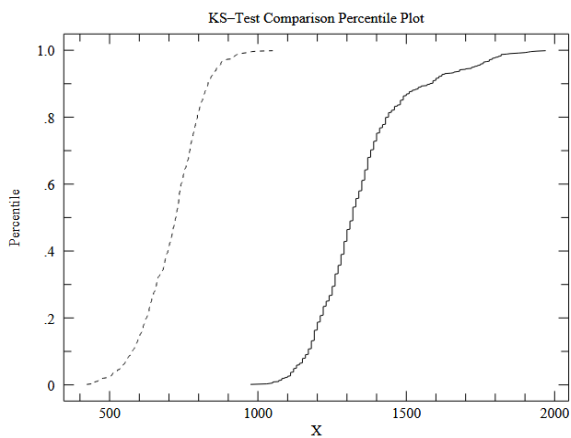

Figure 7. Kolmogorov-Smirnov test[25]. The two cumulative distributions, solid curve is for on-source and the dashed curve for off-source.

We have also used the well known Li-Ma statistics [26], widely used in astronomy and astrophysics for on-source off-source comparisons. Using equation 17 of the Li-Ma paper a very high statistical significance S for the alternative hypothesis that there is a source of light on top of a background is obtained, S =14 [27].

## 4. Conclusions

We have shown in this contribution that there is a real possibility that the NIR scintillation from liquid noble gases may become a viable alternative to the standard VUV light signal in detectors for DM and neutrino physics. The use of the NIR light would avoid the well known problems and difficulties associated with the use of VUV scintillation. We advocate a strong and well funded research program for the study of NIR scintillation in LNGs, one that provides answers to urgent questions such as the accurate determination of the light yield, spectral and time structure of the signal and finally the development of photon detection systems using the NIR light in LNG TPCs systems.

## Acknowledgments

The technical staff at the Fermilab Proton Assembly Building is thanked for their support. Fermilab is Operated by Fermi Research Alliance, LLC under Contract No. De-AC02-07CH11359 with the United States Department of Energy.

$N_{Off}$ = 712, and for the ratio of (on-source)/(off-source) observation times (called $\alpha$ in Li-Ma's paper), $\alpha$ =1. The same significance S would be obtained had we used the usual expression

$S = (N_{on} - N_{off})/\sqrt{N_{off}}$ .